\documentclass[11pt]{article}
\usepackage{a4wide}
\usepackage{epsfig,epsf,graphics}
\usepackage{times}
\usepackage{color}
\usepackage{amsfonts,amssymb,latexsym,amsmath} 
\usepackage{bm}
\usepackage{enumerate}
\usepackage{enumitem}
\usepackage{hhline}
\usepackage{longtable}
\usepackage{array} 
\usepackage{epstopdf}

\newcommand{\partiallr}{\overleftrightarrow{\partial}}
\newcommand{\al}{&\!\!\!\!}
\newcommand{\Lag}{\mathcal{L}}

\newcommand{\Tr}{\textrm{Tr}}

\newcommand{\X}{X(3872)}
\newcommand{\Y}{Y(4260)}

\newcommand{\email}[1]{\footnote{{\em E-mail address:} \texttt{#1}}}

\begin{document}
\title{Production of the $X(3872)$ in charmonia radiative decays}

\author{Feng-Kun Guo$^{\, a,}$\email{fkguo@hiskp.uni-bonn.de} ,
Christoph Hanhart$^{\,b,}$\email{c.hanhart@fz-juelich.de} ,
Ulf-G.~Mei{\ss}ner$^{\, a,b,}$\email{meissner@hiskp.uni-bonn.de} ,
Qian Wang$^{\, b,}$\email{q.wang@fz-juelich.de} ,
Qiang Zhao$^{\, c,}$\email{zhaoq@ihep.ac.cn} \\ 
   {\it\small$^a$Helmholtz-Institut f\"ur Strahlen- und Kernphysik and Bethe
   Center for Theoretical Physics, }\\
   {\it\small Universit\"at Bonn,  D-53115 Bonn, Germany }\\
   {\it\small$^b$Institut f\"{u}r Kernphysik, Institute for Advanced 
Simulation, and J\"ulich Center for Hadron
Physics,}\\
   {\it \small D-52425 J\"{u}lich, Germany} \\
   {\it\small$^c$Institute of High Energy Physics and Theoretical Physics Center 
for Science Facilities, } \\ 
   {\it\small  Chinese Academy of Sciences, Beijing 100049, China}
   }

\maketitle

\begin{abstract}
\noindent
We discuss the possibilities of producing the $\X$, which is assumed to be a $ 
D\bar D^*$ bound state, in radiative decays of charmonia. We argue that the ideal
energy regions to observe the $\X$ associated with a photon in 
$e^+e^-$--annihilations are around the $\Y$ mass and around 4.45~GeV, due to the 
presence of the $S$-wave $D\bar D_1(2420)$ and $D^{*}\bar D_1(2420)$ threshold, 
respectively. Especially, if the $\Y$ is dominantly a $D\bar D_1$ molecule and 
the $\X$ a $D\bar D^*$ molecule, the radiative transition strength will be quite 
large.
\end{abstract}

\newpage

\section{Introduction}

Since its discovery by the Belle Collaboration~\cite{Choi:2003ue}, the $\X$,
which is extremely close to the $D^0\bar D^{*0}$ threshold, has stimulated a lot
of efforts, both experimental and theoretical. It is regarded as one of the most
promising candidates for a hadronic molecule, which are formed of two or more
hadrons --- analogous to the deuteron, the shallow bound state made of a proton
and a neutron. The quantum numbers of the $\X$ have been determined
to be $J^{PC}=1^{++}$~\cite{Aaij:2013zoa}, in accordance with the
hadronic molecular interpretations which can be either an $S$-wave
bound state~\cite{rujula,voloshin,Tornqvist:2004qy,swanson} or a virtual state in the  $ D\bar D^*$
system~\cite{Hanhart:2007yq}.
Another puzzling new charmonium state is the $\Y$ with quantum numbers
$J^{PC}=1^{--}$, which was observed by the
BaBar Collaboration~\cite{Aubert:2005rm}. It is difficult to be put in the
vector family of the $ c\bar c$ in potential models. Various interpretations
were proposed. One intriguing possibility is that the main component of the $\Y$
is a $D\bar D_1(2420)$ bound
state~\cite{Ding:2008gr,Li:2013bca,Wang:2013cya}.~\footnote{Notice that there
are two $D_1$  states of similar masses, and the one in question should be the
narrower one, i.e. the $D_1(2420)$, because it is not sensible to discuss a
constituent with a width comparable or even larger than the range of
forces~\cite{Filin:2010se,Guo:2011dd}.} For a comprehensive review of the
$\X$, $\Y$ and other $XYZ$ states observed in the last decade, we refer to
Ref.~\cite{Brambilla:2010cs}.

So far the $\X$ has been observed in several different processes. The discovery 
was made in $B$-meson decays in the processes $B^\pm\to K^\pm J/\psi \pi^+\pi^-$ 
by the Belle Collaboration ~\cite{Choi:2003ue} and later confirmed by the BaBar 
Collaboration~\cite{Aubert:2004ns}. It was also observed in the 
proton--antiproton annihilations  $p\bar{p}\to J/\psi \pi^+\pi^- X$ by both the 
CDF~\cite{Acosta:2003zx} and D0~\cite{Abazov:2004kp} Collaborations, and in 
proton--proton collisions by the LHCb 
Collaboration~\cite{Aaij:2013zoa,Aaij:2011sn}. It is quite natural to search for 
the $\X$ also in the decays of higher charmonia, especially  the $1^{--}$ 
states, which can be easily and copiously produced in electron-positron 
collisions at, e.g., the Beijing Electron-Positron Collider II (BEPC-II). However,
so far no evidence of the $\X$ in the radiative charmonium decays has been 
reported. In this paper, we will investigate the production of the $\X$ in
the radiative decays of charmonium states, which include the $ \psi(4040)$, the
$\psi(4160)$, the $\Y$ and the $\psi(4415)$, which are all in the energy range
of the BESIII experiment~\cite{Asner:2008nq} at the BEPC-II. As will be shown later on, among
the vector charmonium(-like) states, the $\Y$ is the most promising one for
producing the $\X$, if
the long-distance part of its wave function is dominated by the $D\bar
D_1$ hadronic molecule component ---  note that the mass of the $\Y$  is located close to the $S$-wave $D\bar D_1$ threshold.

Our paper is organized as follows: In Sec.~\ref{sec:pc}, based on a
nonrelativistic effective field theory (NREFT), we will identify the most
important mechanism for the $\X$ production, namely the triangle loops with the
coupling of the initial charmonium(-like) state with charmed mesons being
$S$-wave. Using the effective Lagrangians given in Sec.~\ref{sec:lag}, we
will calculate the partial decay widths of the radiative transitions of the
charmonia, especially parameter-free predictions for the $ \Y \to \X \gamma$ 
and will be made, and the results will be given in Sec.~\ref{sec:res}. A
brief summary will be given in the last section.

\section{Identifying the most important mechanism}
\label{sec:pc}

In general, a hadronic molecule is not a pure two-meson state  since it can couple to other
components, such as a $q\bar q$  or a compact multiquark state, when these
have the same quantum numbers. Thus, such a hadronic molecule can be produced
through either the compact quark component or the hadronic constituents. It is
a process-dependent question and in some cases one of those two mechanisms is more
important than the other. Let us take the $\X$ as an example, which may be decomposed as
\begin{equation}
    |\X \rangle = \alpha_1 | c\bar c \rangle + \frac{ \alpha_2}{ \sqrt{2} }
\left| D\bar D^{*} + c.c. \right\rangle~.
\end{equation}
Then the production amplitude is composed of two parts, $ \mathcal{P}_{\X} =
\alpha_1\, \mathcal{P}_{c\bar c} + \alpha_2\, \mathcal{P}_{ D\bar D^*}  $, where 
$\mathcal{P}_{ c\bar c}$ and $ \mathcal{P}_{ D\bar D^*}$ represents the
production of the $c\bar c$ and $D\bar D^*+c.c.$, respectively (in the
following, the charge conjugated channel will not be shown for simplicity but
will be included in the numerical calculation). We assume that the $\X$ is
mainly a $D\bar D^*$ molecule, i.e. $| \alpha_2|\gg |\alpha_1 | $. In this case, 
if $\mathcal{P}_{ D\bar D^*}$ is not heavily  suppressed,
then the $\X$ will dominantly be
produced through the long distance $D\bar D^*$ component --- see Refs.~\cite{erics,interplay}
for a more detailed discussion.

\begin{figure}[tb]
\begin{center}
  \includegraphics[width=\textwidth]{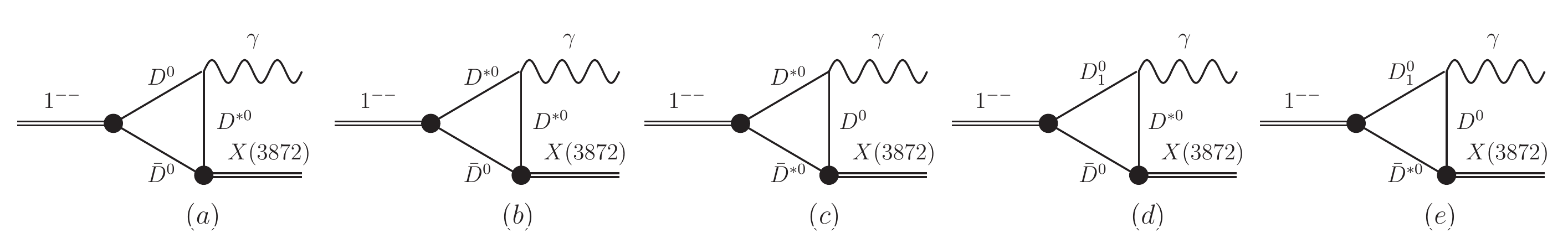}
\caption{Relevant triangle loops for the production of the $X(3872)$ in the
vector charmonium radiative decays. The charge-conjugated diagrams are not shown.
}
\label{fig:FeynmanDiagram}
\end{center}
\end{figure}
Both the short and long distance production of the $\X$ in the radiative decays
of the $\psi(4040)$ and $ \psi(4160)$ are considered in
Refs.~\cite{Mehen:2011ds,Margaryan:2013tta} in the framework of the so-called
X-EFT~\cite{Fleming:2007rp}. Here, we will focus on the contribution from
intermediate charmed meson loops, i.e. the quantity $ \mathcal{P}_{ D\bar D^*}$
defined above. The mechanism is shown in Fig.~\ref{fig:FeynmanDiagram}.
Both the initial charmonium and the $\X$ couple to a pair of charmed and
anticharmed mesons. The $\X$ couples to the $D\bar D^*$ pair in an $S$-wave.
With the quantum numbers being $1^{--}$, the initial charmonium can couple to
either two $S$-wave charmed mesons in a $P$-wave, or one $P$-wave and one
$S$-wave charmed mesons in an $S$- or $D$-wave. As we will show in the
following, the
mechanism with an $S$-wave coupling to the initial charmonium will greatly
facilitate the production processes.

The charmed meson channels which could have significant effects are those close
to the mass of the considered charmonium. In this work we will consider the
$s_\ell^P=\frac12^-$ ($S$-wave), and $s_\ell^P=\frac32^+$ ($P$-wave) charmed 
mesons, where $s_\ell$ is the total angular momentum of
the light quark system which includes the light quark spin and orbital angular
momentum. The pertinent triangle loops are shown in
Fig.~\ref{fig:FeynmanDiagram}. There is no $D_2$ analogue of 
diagrams (d,e) because, different from the $ D_1$ case, its quantum numbers 
does not allow that all vertices are in $S$-wave. One should notice that 
although the $\X$ can have a sizable $ D^+D^{*-}$ 
component~\cite{Gamermann:2009fv,Gamermann:2009uq}, because the magnetic 
coupling to the neutral charmed mesons is much larger than that to the charged 
ones, see, e.g. Ref.~\cite{Hu:2005gf}, we only consider the neutral charmed 
mesons in the loops.~\footnote{In fact, there can be photonic coupling to the 
charged charmed mesons from gauging the $\psi D^{(*)}\bar D^{(*)}$ vertex and 
the kinetic energy of the charmed mesons, see e.g.~\cite{Mehen:2011tp}. 
However, they are of order $ \mathcal{O}{(v)}$, thus less important, in the 
power counting scheme to be detailed in the following. Furthermore, the loops 
involving such vertices are divergent and hence need unkown counterterms. In 
contrast, Ref.~\cite{Aceti:2012cb} states that including the charged charmed 
mesons would largely increase the partial decay width of the $\X\to\gamma 
J/\psi$ based on a vector meson dominance model in a flavor SU(4) formalism.}

Because all the charmonia considered are close to the open charm thresholds in
question, the intermediate charmed and anticharmed mesons are nonrelativistic.
We are thus allowed to use a nonrelativistic power counting, the framework of
which was introduced for studying the intermediate meson loop effects in
certain hadronic transitions of charmonia  in
Refs.~\cite{Guo:2009wr,Guo:2010zk,Guo:2010ak}. Being nonrelativistic, the
velocity of the intermediate mesons $v$ is much smaller than 1. Thus,  the loop
diagrams as shown in Fig.~\ref{fig:FeynmanDiagram} can be organized
through a velocity counting, where the
three-momentum scales as $v$, the kinetic energy scales as $v^2$, and each of
the nonrelativistic propagators scales as $v^{-2}$. In leading order,
 the $S$-wave coupling is momentum independent and does not
contribute any power to the velocity counting. The $P$-wave coupling scales
as $v$~\cite{Guo:2009wr} or as the external momentum~\cite{Guo:2010zk,Guo:2010ak}
depending on the process in question.

Let us focus on the last two diagrams of Fig.~\ref{fig:FeynmanDiagram} first.  
The $D$ meson has $s_\ell=1/2$, and $D_1$ has $s_\ell=3/2$. Thus, they can 
couple to $L=2$ but not to $L=0$, where $L$ is the orbital angular momentum, in 
the heavy quark limit. As a result, only the $D$-wave charmonia can couple to 
the $D\bar D_1$ in an $S$-wave, and for the $S$-wave charmonia the coupling 
must be $D$-wave. Thus, if the initial state is a $D$-wave charmonium or  has a 
significant $D\bar D_1$ molecular component (as might be the case for the $\Y$), 
the loop integral scales as
\begin{equation}
\label{eq:SSpc}
\frac{v^5}{(v^2)^3} E_\gamma = \frac{E_ \gamma}{v},
\end{equation}
where $E_\gamma$ is the external photon energy. The decay amplitude is the
product of the loop integral and the coupling constants for the three vertices.
One sees that the amplitude is greatly enhanced for small velocity. It was
shown in Ref.~\cite{Guo:2012tg} that the value of the velocity should be
understood as the average of two velocities which correspond to the two cuts in
the triangle diagram. These two velocities may be estimated as $\sqrt{| m_1 +
m_2 - M_i |/\bar m_{12}}$ and  $\sqrt{| m_2 +
m_3 - M_f |/\bar m_{23}}$, where $m_{2}$ is the mass of the charmed meson
between the two charmonia, $m_{1(3)}$ is the mass of the meson between the
initial (final) charmonium and the photon, $\bar m_{ij}=(m_i+m_j)/2$, and
$M_{i(f)}$ is the mass of the initial (final) charmonium. Therefore, the
amplitude is most enhanced when both the initial and final charmonia are close
to the corresponding thresholds.

For diagrams (a), (b) and (c) of Fig.~\ref{fig:FeynmanDiagram},  the vertex
involving the initial charmonium is in a $P$-wave. The momentum in that vertex has to be
contracted with the external photon momentum $q$, and thus should be counted as
$q$. The decay amplitude through this type of loops scales as
\begin{equation}
    \label{eq:PSpc}
\frac{v^5}{(v^2)^3} \frac{q^2}{m_0} = \frac{E_\gamma^2}{m_0\, v},
\end{equation}
where $ m_0$ is a quantity of the dimension mass, and the factor of $m_0^{-1}$
is introduced to make the above expression have the same dimension as that
obtained in Eq.~\eqref{eq:SSpc}. This factor in fact accounts for the different
dimensions of the coupling constants for the $P$-wave and $S$-wave vertices in
diagrams (a, b, c) and (d, e), respectively, i.e. $m_0 = |g_4/g_3|$ where $g_3$ 
and
$g_4$ are the coupling constants to be defined in Eq.~\eqref{eq:lagD} below. If all
the coupling constants are of natural size, that
is $m_0\sim 1$~GeV, then this loop should be suppressed relative to the one
in Eq.~\eqref{eq:SSpc} for a soft photon. This is supported by the numerical
results in Sec.~\ref{sec:res}. Notice that only neutral charmed mesons are
involved so that the $P$-wave vertex, although it contains a derivative, will 
not get gauged 
and the triangle 
diagrams are gauge invariant.

If the initial charmonium is the $\psi(4040)$ or the $ \psi(4415)$, which are
the radial exceptions of $J/\psi$ and thus $S$-wave charmonia, the
coupling to the $D\bar D_1$ is in a $D$-wave in the heavy quark limit, as outlined above. In this
case, the $\psi D\bar D_1$ vertex should be counted as $v^2$. Using the same
power counting, the loops in Fig.~\ref{fig:FeynmanDiagram} (d, e) should scale
as $v\,E_\gamma$, and thus are suppressed rather than enhanced
for small values of $v$. 

In the above discussions, we have neglected the width of the $D_1(2420)$, which
presents a new scale. One concern is whether it would break the power
counting established above. The width of the $ D_1(2420)$ is
$27.1\pm2.7$~MeV~\cite{Beringer:1900zz}, thus $ \Gamma_1 \lesssim |2\,
b_{12}|$, where $b_{12}=m_1+m_2-M_i$. From Eq.~\eqref{eq:nrloop_width}, which
is the nonrelativistic scalar loop function where one of the intermediate mesons
carries a finite  constant width, one can conclude that the power counting
scheme will not be modified by the presence of the finite width of the
$D_1(2420)$ (as long as the width is sufficiently small).

\section{Effective Lagrangians}
\label{sec:lag}

Because the charmed mesons do not have definite charge parity, it is necessary 
to clarify the phase convention under charge conjugation to be used in our 
paper, which is
\begin{equation}
\label{eq:charge}
  \mathcal{C} D \mathcal{C}^{-1} = \bar D,\quad
  \mathcal{C} D^* \mathcal{C}^{-1} = \bar D^*,\quad
  \mathcal{C} D_1 \mathcal{C}^{-1} = \bar D_1.
\end{equation}
The $\X$ has a positive $C$-parity, and the $\Y$ as well as all the other 
vector charmonium states have  negative $C$-parity. Thus, the flavor wave 
functions of the $\X$ and $\Y$ in terms of the charmed mesons are convention 
dependent. With the convention specified above,
the $D\bar D^*$ and $D\bar D_1$ components of the $\X$ and 
$\Y$ can be written
as~\footnote{In the literature, some authors write the wave function of the
$\X$ with a different relative sign of the two terms, $|\X\rangle =
\frac1{\sqrt{2}} \left|D\bar D^* - D^*\bar D\right\rangle$. This corresponds to
a different convention for the $C$-parity transformation for the $D^*$,
$\mathcal{C} D^* \mathcal{C}^{-1} = - \bar D^*$. Notice that only the flavor
neutral mesons are eigenstates of the $C$-parity, the physical observables
should be independent of the convention. For a detailed discussion in the
case of the $\X$, see Ref.~\cite{Thomas:2008ja}.}
\begin{equation}
  |\X\rangle = \frac1{\sqrt{2}} \left|D\bar D^* + D^*\bar D\right\rangle, \quad
  |\Y\rangle = \frac1{\sqrt{2}} \left|D_1\bar D - D\bar D_1\right\rangle.
\end{equation}

Because we work with nonrelativistic kinematics for the charmed mesons
and charmonia throughout this work, the two-component notation introduced in
Ref.~\cite{Hu:2005gf} is very convenient. In this simplified notation, the
field for the ground state charmed mesons is $H_a = \vec{V}_a\cdot \vec{\sigma}
+ P_a$, where $\vec\sigma$ are the Pauli matrices, $P_a$ and $V_a$ annihilates
the pseudoscalar and vector charmed mesons, respectively, and $a$ is the flavor
label for the light quarks. The quantum numbers of the light quark system in
these two mesons are $s_\ell^P=\frac12^-$. Under the convention specified in
Eq.~\eqref{eq:charge}, the field annihilating the ground state mesons
containing an anticharm quark is~\cite{Fleming:2008yn}
\begin{equation}
  \bar H_a = \sigma_2 \left( \vec{\bar V}_a\cdot \vec{\sigma}^T + \bar P_a
\right) \sigma_2 = -\vec{\bar V}_a\cdot \vec{\sigma} + \bar P_a.
\end{equation}
The field for the $s_\ell^P=3/2^+$ charmed mesons can be written as
\begin{equation}
  T_a^i = P_{2a}^{ij} \sigma^j + \sqrt{\frac23}\, P_{1a}^i + i \sqrt{\frac16}\,
\epsilon_{ijk} P_{1a}^j \sigma^k,
\end{equation}
where $P_{1a}$ and $P_{2a}$ annihilate the charmed mesons $D_1(2420)$ and
$D_2(2460)$, respectively. The charmed antimesons are collected in  $ \bar
T_a^i = -\bar P_{2a}^{ij} \sigma^j + \sqrt{2/3}\, \bar
P_{1a}^i - i \sqrt{1/6}\, \epsilon_{ijk} \bar P_{1a}^j \sigma^k $, where the
convention $\mathcal{C} D_2 \mathcal{C}^{-1} = \bar D_2$ is adopted.
Under parity and charge conjugation and with the convention specified above,
these fields transform as
\begin{eqnarray}
    H_a \al\overset{ \mathcal{P}}{\rightarrow}\al - H_a, \qquad   H_a \overset{
\mathcal{C}}{\rightarrow} \sigma_2 \bar H_a^T \sigma_2, \qquad
    \bar H_a \overset{ \mathcal{P}}{\rightarrow} - \bar H_a, \qquad  \bar
H_a \overset{ \mathcal{C}}{\rightarrow} \sigma_2 H_a^T \sigma_2~,\\
T_a^i \al\overset{ \mathcal{P}}{\rightarrow}\al T_a^i, \qquad \quad~
T_a^i \overset{
\mathcal{C}}{\rightarrow} \sigma_2 \bar T_a^{i\,T}\sigma_2, \qquad
    \bar T_a^i\overset{ \mathcal{P}}{\rightarrow} \bar T_a^i, \qquad\quad~
\bar T_a^i \overset{ \mathcal{C}}{\rightarrow} \sigma_2 T_a^{i\,T} \sigma_2~.
\end{eqnarray}

Analogously, we can construct the field for the $S$-wave charmonia, which is
$J = \vec{ \psi}\cdot \vec{\sigma} + \eta_c $, where $ \psi$ and $ \eta_c$
annihilate the vector and pseudoscalar charmonia, respectively. The leading
coupling of the $S$-wave charmonium with the charmed and anticharmed mesons
reads as
\begin{equation}
  \label{leq:lagS}
    \mathcal{L}_{S} = i \frac{g_2}{2} \left \langle  \bar{H}_a^\dag \, \vec{
\sigma}
\cdot \partiallr H_a^\dag \, J \right \rangle 
+ \text{H.c.},
\end{equation}
where $A\overleftrightarrow{\partial} B\equiv A(\vec{\partial}B) -
(\vec{\partial}A)B$ and $\langle \ldots \rangle$ denotes the 
trace in flavor space. Notice that all the charmed meson and charmonium fields in
the above Lagrangian and the ones in the following are nonrelativistic and have
dimension  mass$^{3/2}$.

Some of the $1^{--}$ charmonia in question are $D$-wave
states. For instance, the $\psi(4160)$ is widely considered as the $2\,^3D_1$
state~\cite{Godfrey:1985xj,Barnes:2005pb}. The field for the $D$-wave charmonia 
in
two-component notation can be written as~\cite{Margaryan:2013tta}
\begin{equation}
  J^{ij} = \frac12 \sqrt{\frac35} \left( \psi^i\sigma^j + \psi^j \sigma^i
\right) - \frac{1}{ \sqrt{15}} \delta^{ij} \,\vec \psi\cdot \vec\sigma ,
\end{equation}
where only the $1^{--}$ state relevant for our discussions is included.
Considering parity, $C$-parity, spin symmetry and Galilean invariance, the
leading order Lagrangian for the coupling of the $D$-wave charmonia to the
charmed and anticharmed mesons can be written as
\begin{equation}
  \label{eq:lagD}
  \mathcal{L}_{D} = i \frac{g_3}{2} \left \langle \bar{H}_a^\dag \, \sigma^i
\partiallr^j H_a^\dag \, J^{ij} \right \rangle + \frac{g_4}{2} \left \langle
\left( \bar{T}^{j\,\dag}\, \sigma^i H^\dag - \bar{H}^\dag \,\sigma^i
T^{j\,\dag} \right) J^{ij} \right \rangle + \text{H.c.},
\end{equation}
where the first term has already been introduced in
Ref.~\cite{Margaryan:2013tta} ($g_3$ is denoted by $g$ in that paper).

In order to calculate the triangle diagrams depicted in
Fig.~\ref{fig:FeynmanDiagram}, we need to know the photonic coupling to the
charmed mesons. The magnetic coupling of the photon to the $S$-wave heavy mesons
is described by the Lagrangian~\cite{Amundson:1992yp,Hu:2005gf}
\begin{equation}
\label{eq:Lem} \Lag_{HH\gamma} = \frac{e\,\beta}{2} \Tr\left[ H_a^\dag H_b\,
\vec{\sigma}\cdot \vec{B} \, Q_{ab} \right] + \frac{e\, Q'}{2m_Q} \Tr \left[
H_a^\dag \,
\vec{\sigma}\cdot \vec{B} \, H_a\right],
\end{equation}
where $B^k=\epsilon^{ijk}\partial^iA^j$ is the magnetic field, $Q$ is the light
quark charge matrix, and $Q'$ is the heavy quark electric charge (in units of 
the proton charge $e$). These two terms describe the magnetic coupling due to 
the light and heavy quarks, respectively. The E1 transition of the $\frac32^+$ 
charmed mesons to the $\frac12^-$ states may be parameterized in terms of a 
simple Lagrangian
\begin{equation}
  \label{eq:LTHga}
  \Lag_{TH\gamma} = \sum_a\frac{c_a}{2}\, \Tr \left[ T_a^i H_a^\dag \right] E^i
+ \text{H.c.}
\end{equation}
Note that here the coefficients are light-flavor-dependent.

At last, assuming that the $\X$ and $\Y$ are hadronic molecules, we parameterize 
their coupling to the charmed mesons in terms of the following Lagrangian
\begin{equation}
  \label{eq:LXY}
  \Lag_{XY} = \frac{y}{\sqrt{2}} Y^{i\,\dag} \left( D_{1a}^i \bar D_a - D_a
\bar D_{1a}^i \right) + \frac{x}{ \sqrt{2} } X^{i\,\dag} \left( D^{*0\,i} \bar 
D^0 +
D^0 \bar D^{*0\,i} \right) + \text{H.c.},
\end{equation}
where we assume that the $\Y$ couples to the $ D\bar D_1$ in an isospin
symmetric manner so that the light flavor index $a$ runs through $u$ and $d$,
and neglect all the other components except for the $ D^0\bar D^{*0}$ for the
$\X$.

\section{Results and discussion}
\label{sec:res}

Considering a state slightly below an $S$-wave two-hadron threshold, the
effective coupling of this state to the two-body channel is related to the
probability of finding the two-hadron component in the physical wave function of 
the bound state, $\lambda^2$,
and the binding energy, $ \epsilon=m_1+m_2-M$~\cite{Weinberg:1965zz,Baru:2003qq}
\begin{equation}
  \label{eq:gbs}
  g_\text{NR}^2 = \lambda^2 \frac{16\pi}{\mu}
\sqrt{\frac{2\epsilon}{\mu}}
\left[ 1 + \mathcal{O} \left( \sqrt{2\mu \epsilon}\,r \right) \right],
\end{equation}
where $\mu=m_1 m_2/(m_1+m_2)$ is the reduced mass and $r$ is the range of 
forces, and the
nonrelativistic normalization is used. After proper
renormalization (see Ref.~\cite{Cleven:2011gp}), the coupling
constants in Eq.~\eqref{eq:LXY} are given by the one in the above equation. 
Notice that the coupling constant gets maximized for a pure bound state, which 
has $\lambda^2=1$ by definition.

The threshold of the $D^0$ and $D^{*0}$ using
the PDG fit values for the masses~\cite{Beringer:1900zz} is
$3871.84\pm0.20$~MeV. The mass of the $\X$ is
$3871.68\pm0.17$~MeV~\cite{Beringer:1900zz}. With $M_Y=4263^{+8}_{-9}$~MeV, and
the isospin averaged masses of the $D$ and $D_1$ mesons, we obtain the mass differences
between the $\X$ and $\Y$ and their corresponding thresholds, respectively,
\begin{equation}
   M_{D^0}+ M_{D^{*0}} - M_X = 0.16\pm 0.26~\text{MeV}, \qquad  M_{D}+
M_{D_1(2420)} - M_Y = 27^{+9}_{-8}~\text{MeV}.
\end{equation}
Assuming that the $\X$ and $\Y$ are pure hadronic molecules, which corresponds
to the probability of finding the physical states in the two-hadron states 
$\lambda^2 = 1$, we obtain
\begin{equation}
  |x| = 0.97^{+0.40}_{-0.97}\pm0.14~\text{GeV}^{-1/2} , \qquad |y| =
3.28^{+0.25}_{-0.28}\pm1.39~\text{GeV}^{-1/2}~,
\end{equation}
where the first errors are from the uncertainties of the binding energies, and
the second ones are due to the approximate nature of Eq.~\eqref{eq:gbs}. The
range of forces is estimated by $r^{-1}\sim \sqrt{2\, \mu\, \Delta_\text{th}}$
where $\mu$ is the reduced mass and $ \Delta_\text{th}$ is the difference
between the threshold of the components and the next close one, which is
$M_{D^{*+}} + M_{ D^+ } - M_{D^{*0}} - M_{ D^0 }$ for the $\X$ and $M_{D_1} +
M_{ D^* } - M_{D_1} - M_{ D }$ for the $\Y$, respectively.

The  value of $\beta$ in the magnetic coupling of the $S$-wave charmed
mesons is not precisely known. We  will use the value
$\beta^{-1}=276$~MeV determined with $m_c=1.5$~GeV in Ref.~\cite{Hu:2005gf}.
There is no experimental measurement on the radiative decays of the $P$-wave
charmed mesons. However, there have been a few calculations using various
quark models. Taking the predictions of $\Gamma( D_1^0 \to D^{(*)0}\gamma)$ in
Refs.~\cite{Fayyazuddin:1994qu,Korner:1992pz,Godfrey:2005ww} as a guidance, the
value for the coupling constant for the neutral charmed mesons $c_0$ is in the
range $[0.3,0.5]$.

\subsection{$\bm{\psi(4040)\to\gamma X(3872)}$ and $\bm{\psi(4415)\to\gamma 
X(3872)}$}
\label{sec:4040}

The $ \psi(4040)$ and $ \psi(4415)$ were widely accepted as the $3S$ and $4S$
vector charmonium states, respectively~\cite{Godfrey:1985xj}. In the heavy
quark limit, spin symmetry requires that the $S$-wave charmonium couples to the
$D^{(*)} \bar D_1$ in a $D$-wave. As shown in Sec.~\ref{sec:pc}, such a $D$-wave
vertex will cause the charmed meson loops to be suppressed. Thus, we will neglect
these loops, and consider only the loops involving the $S$-wave charmed mesons
$D$ and $D^*$, which correspond to the diagrams shown in
Fig.~\ref{fig:FeynmanDiagram} (a),  (b) and (c). Assuming that the two-body $S$-wave
charmed mesons saturate the decay width of the $ \psi(4040)$ and 90\% of
width of the $ \psi (4415)$ --- the only relatively well measured branching
fraction is the sequential decay into the $D^0 D^- \pi^+ + c.c.$ through the $D
\bar D_2(2460)$ which is $(10\pm4)\%$, we may obtain an upper limit for the
coupling constant $g_2$ for both the $3S$ and $4S$ charmonium states,
\begin{equation}
  \left|g_{2[3S]}\right| < 0.85~\text{GeV}^{-3/2}~, \qquad
\left|g_{2[4S]}\right| < 0.23~\text{GeV}^{-3/2} \ .
\end{equation}
As a result, the upper limits for the production of the $\X$ are
\begin{equation}
  \Gamma( \psi(4040)\to \gamma \X )_\text{(a,b,c)} < 0.25~\text{keV}, \qquad 
\Gamma(
\psi(4415)\to \gamma \X )_\text{(a,b,c)} < 0.63~\text{keV},
\end{equation}
which correspond to tiny branching fractions of order $10^{-5}$.

However, even a small $D$-wave $ c\bar c$ mixture would greatly enhance the
decay width of the $ \psi(4415)$. This is because the $ \psi(4415)$ is
only 10~MeV below the $ D^*\bar D_1$ threshold, and the velocity, the relevant parameter for the power
counting, is as small as 0.04. Considering such an admixture, we obtain from the 
last two diagrams in Fig.~\ref{fig:FeynmanDiagram}
\begin{equation}
  \label{eq:wid4415}
  \Gamma( \psi(4415)\to \gamma \X)_\text{(d,e)} = 287 \sin^2\theta\left(g_4\, x
\,\text{GeV} \right)^2 c_0^2 ~\text{keV} \lesssim 89 \sin^2\theta
\left(g_4^2\,\text{GeV}
\right) ~\text{keV}~,
\end{equation}
where $c_0\simeq0.4$ is used, and $\sin\theta$ is the mixture of the $D$-wave
component in the $ \psi(4415)$ wave function. In Ref.~\cite{Badalian:2008dv}, 
$\theta\approx 34^\circ$ is suggested from an analysis of the $e^+e^-$ decay 
widths of the vector charmonia. We have assumed spin symmetry for the coupling 
of the initial charmonium to the charmed mesons.

\subsection{$\bm{ \psi(4160)\to\gamma X(3872)}$}
\label{sec:4160}

As discussed before, being the $2D$ charmonium state, the $ \psi(4160)$ couples
to a pair of $S$-wave charmed mesons in a $P$-wave, and to one $S$-wave and one
$P$-wave charmed mesons in an $S$-wave. Thus all the diagrams shown in
Fig.~\ref{fig:FeynmanDiagram} contribute to its radiative decay into the $\X$.
For the diagrams (a), (b) and (c), we can derive an upper limit for their
contributions. The upper limit for the coupling $g_3$ for the $ \psi(4160)$ may
be obtained by saturating its total decay width by two-body decays into a pair of
$S$-wave charmed mesons. We obtain $\left|g_{3[2D]}\right|<0.72$~GeV$^{-3/2}$.
Using this value, the contribution
of the $S$-wave charmed mesons to the width of the $\psi(4160)\to\gamma
X(3872)$ is less than 0.20~keV.  We should mention that our numerical result for 
the width of the $ \psi(4160)\to \gamma \X$ is smaller than 
the estimate in Ref.~\cite{Margaryan:2013tta} 
using a different method and using the BaBar measurement of the 
$\X\to\gamma\psi'$~\cite{Aubert:2008ae}, which was not confirmed by the Belle 
Collaboration~\cite{Bhardwaj:2011dj}, as input.

\begin{figure}[tbh]
    \begin{center}
        \includegraphics[width=0.6\textwidth]{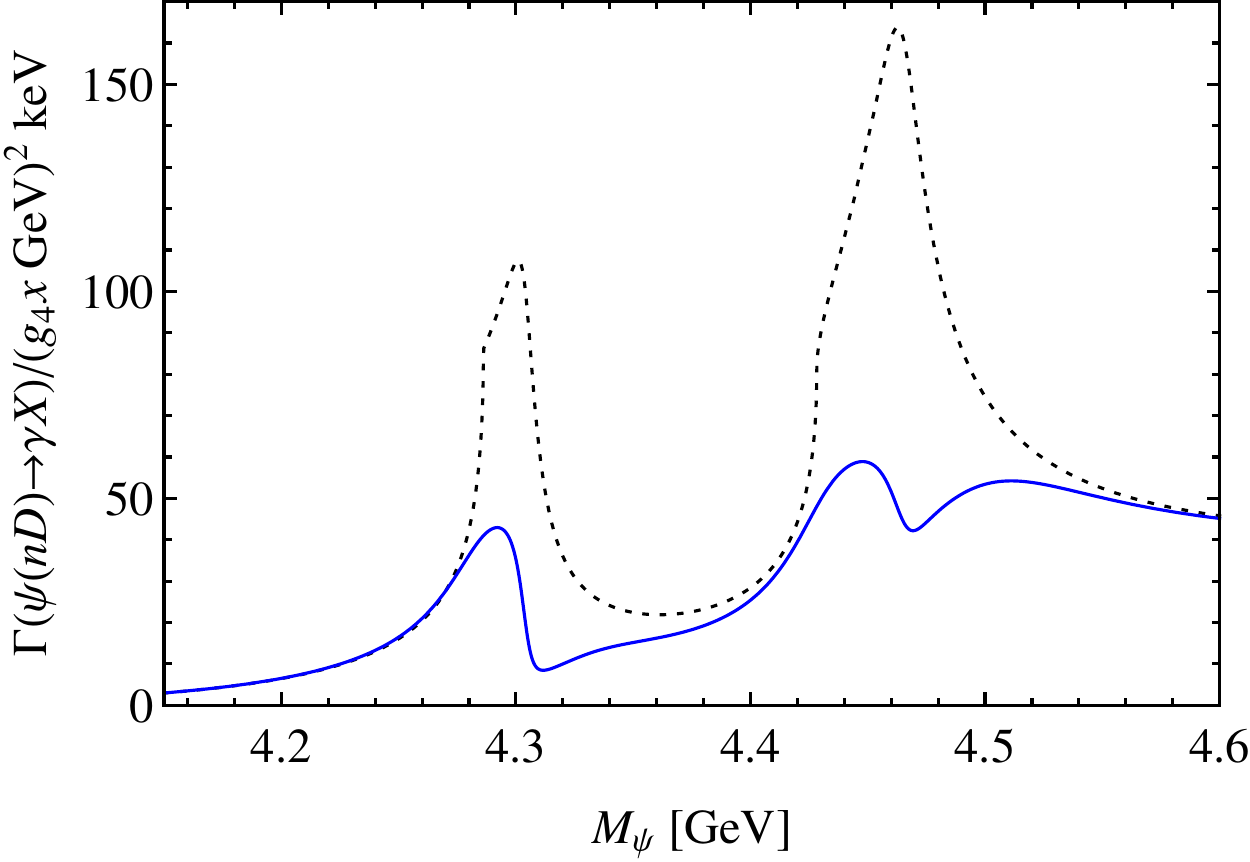}
        \caption{Dependence of the partial decay width of a $D$-wave charmonium 
into 
         $\gamma\X$ on the mass of the 
charmonium. The solid and dotted curves are obtained with and without taking 
into account the width of the $D_1(2420)$, respectively. Here, only the 
contributions from Fig.~\ref{fig:FeynmanDiagram} (d) 
and (e) are included, and $c_0=0.4$ is used.}
\label{fig:psiDxga}
    \end{center}
\end{figure}
The value of $g_4$, which is needed for evaluating the diagrams (d) and (e), is 
unknown. Thus, we express the result from these two diagrams in terms of $g_4$
\begin{equation}
  \label{eq:wid4160}
  \Gamma( \psi(4160)\to \gamma \X)_\text{(d,e)} = 19.4 \left(g_4\, x
\,\text{GeV} \right)^2 c_0^2 ~\text{keV} \lesssim 6.0 \left(g_4^2\,\text{GeV}
\right) ~\text{keV},
\end{equation}
where we have taken $c_0\simeq0.4$. Expressing $g_4$ by $g_4=g_3 m_0$, if
$m_0\sim 1$~GeV, then the approximate upper limit obtained from diagrams (d)
and (e), 3~keV, is one order of magnitude larger than that from diagrams (a),
(b) and (c). This can be understood from the power counting. The momentum of the
photon in this decay is 280~MeV. Thus, the factor $q/m_0$ presents a suppression
of the first three diagrams relative to the last two at the amplitude level.
With the total width of the $\psi(4160)$ being
$103\pm8$`MeV~\cite{Beringer:1900zz}, 
a width of a few keV only amounts to a branching 
fraction of the order of $10^{-5}$. Although larger than the 0.2~keV 
arising from the 
first three diagrams, it is still small so that an experimental 
observation will be difficult. 

However,  notice that the $ \psi(4160)$ is far off the optimized 
region 
for the observation of the $\X$. This can be seen from Fig.~\ref{fig:psiDxga}, 
which shows the dependence of the radiative decay width of a $D$-wave 
charmonium into the $\gamma\X$ on the charmonium mass, where the solid and 
dotted curves represent the results with and without taking into account the 
finite width of the $D_1(2420)$, respectively. One sees pronounced peaks 
slightly above the $ D\bar D_1$ and $D^* \bar D_1$ thresholds in the dashed
curve. This is due to the closeness of the $\X$ to the $ D\bar D^*$ threshold, 
which makes the kinematics so special that $(c'-c)/(2\sqrt{-a\,c})$ --- $a,c$ and 
$c'$ are defined in Eq.~\eqref{eq:abc} --- is close to 1, and thus produces the 
maxima (recall that the imaginary part of $\arctan(i)$ is infinite, c.f. 
Eq.~\eqref{eq:nrloop_nowidth}). The pronounced peaks get smeared by the finite 
width of the $D_1(2420)$, as can be seen from the solid curve. Still, the 
width divided by $g_4^2$ peaks around 4.29~GeV and 4.45~GeV. Thus, as stated 
in Sec.~\ref{sec:4040}, one might be able to make an observation through a 
$D$-wave admixture in the $ \psi(4415)$.

\subsection{$\bm{Y(4260)\to\gamma X(3872)}$}
\label{sec:4260}

We assume that the $\Y$ is a $ D\bar D_1$ molecule according to the suggestions 
of Refs.~\cite{Ding:2008gr,Li:2013bca,Wang:2013cya}. The production of the
recently observed charged charmonium 
$Z_c(3900)$~\cite{Ablikim:2013mio,Liu:2013dau,Xiao:2013iha} can be understood in 
this interpretation~\cite{Wang:2013cya,Wang:2013hga} if it is a $D\bar D^*$
hadronic molecule~\cite{Wang:2013cya,Guo:2013sya,Cui:2013yva,Wilbring:2013cha,
Zhang:2013aoa}. Radiative decays of the $\Y$ into a pair of charmed mesons was 
studied based on this assumption very recently~\cite{Liu:2013}.
In this picture, the radiative decay of the $\Y$ into the $\X$ will be a
long-distance process, and the dominant decay mechanism is shown in
Fig.~\ref{fig:FeynmanDiagram}~(d). With the the loop function given in the
Appendix, we obtain the width
\begin{equation}
  \label{eq:wid4260}
  \Gamma( \psi(4260)\to \gamma \X)_\text{(d)} = 141^{+136}_{-\phantom{0}91}
\left(x^2
\,\text{GeV} \right) c_0^2 ~\text{keV} ,
\end{equation}
where the uncertainty is dominated by the use of Eq.~\eqref{eq:gbs}, which is
mainly due to neglecting the coupled channel $D^*\bar D_1$ in this case. The
velocity counting is well controlled since $v\simeq 0.06$. Using
Eq.~\eqref{eq:nrloop_width}, we have checked that including a finite constant
width for the $D_1$ only causes a minor change of about 3\%.
\begin{figure}[tbh]
\begin{center}
  \includegraphics[width=0.6\textwidth]{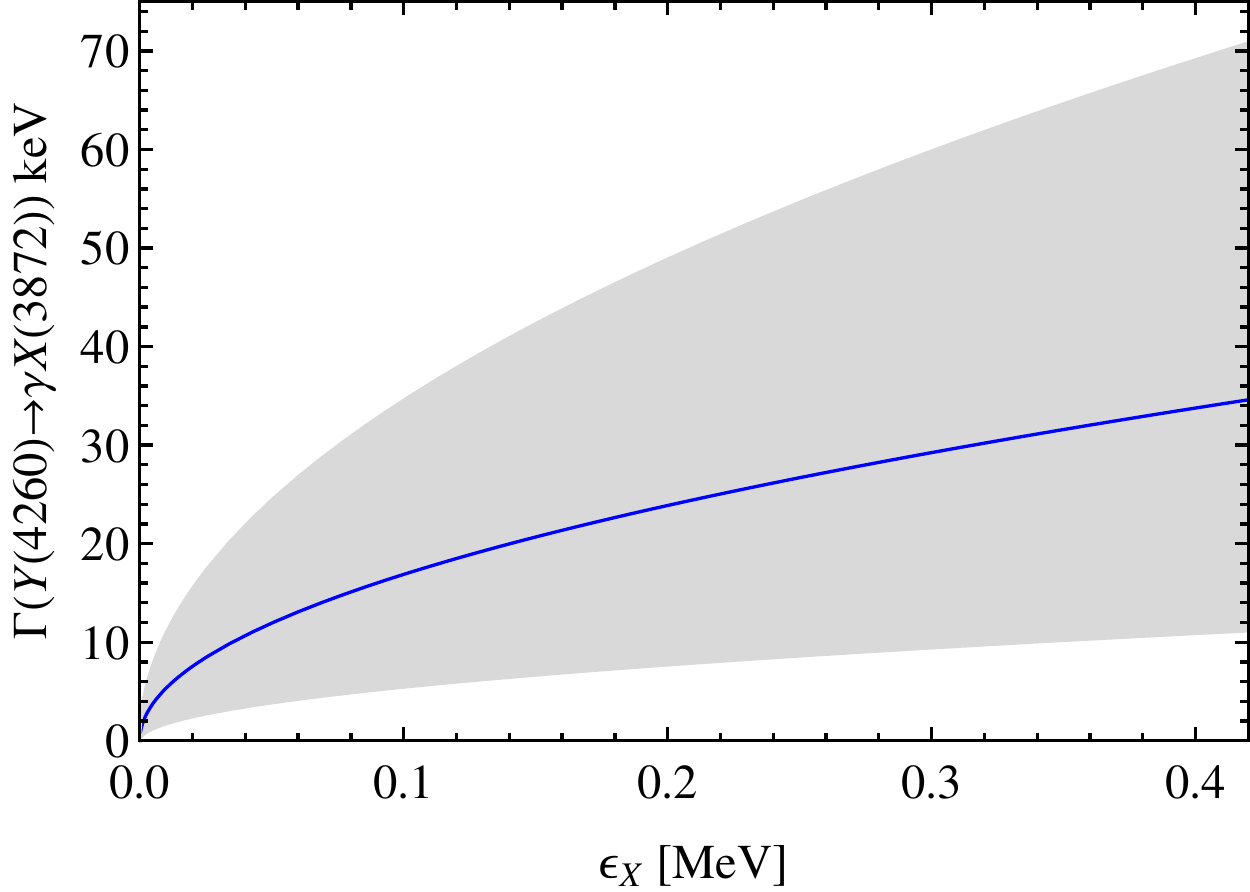}
\caption{Dependence of the width of the $\Y\to\X\gamma$ in terms of the binding
energy of the $\X$, $ \epsilon_X = M_{D^0}+ M_{D^{*0}} - M_X $. Here the $D_1^0
D^{*0}\gamma$ coupling constant is taken as $c_0=0.4$.
}
\label{fig:Ywidth}
\end{center}
\end{figure}
The value of $c_0$ is in the range of $[0.3,0.5]$ using the width predictions
in three different quark
models~\cite{Fayyazuddin:1994qu,Korner:1992pz,Godfrey:2005ww}. Taking $c_0 =
0.4$, we plot the dependence of the width of the $\Y\to \gamma\X$
on the binding energy of the $\X$ in Fig.~\ref{fig:Ywidth}, where the value of 
$x$ is related to the binding energy via Eq.~\eqref{eq:gbs}. Therefore, 
depending on the precise location of the $\X$, the branching fraction can reach 
the order of $10^{-3}$.

\subsection{Using angular distributions to distinguish different loop 
contributions}
\label{sec:ang}

We have argued that the triangle loops with all the intermediate
states being the $S$-wave charmed mesons are suppressed relative to the ones
with one $S$-wave and one $P$-wave charmed mesons when the initial charmonium
is a $D$-wave state. This is based on the assumption that the coupling
constants are of natural size so that $m_0=|g_4/g_3|\sim 1$~GeV. If $g_4$ is
unnaturally small, then these two kinds of mechanisms might be comparable. One 
can check which one is dominant by measuring certain angular distribution. This 
is because the two different types of loops have a different angular dependence 
--- the one with two $S$-wave vertices does not depend on any angle with 
respect to the photon three momentum while the other does, as can be seen from the 
expressions
\begin{eqnarray}
  \label{eq:angA}
   \mathcal{A}_\text{(d,e)} \al=\al A \, (\vec{\epsilon}_\psi \times 
\vec{\epsilon}_\gamma) \cdot \vec{ \epsilon}_X~, \nonumber \\
 \mathcal{A}_\text{(a,b,c)} \al=\al B\,  \hat{q}\cdot\vec{\epsilon}_\psi 
(\hat{q}\times\vec{\epsilon}_X) \cdot \vec{\epsilon}_\gamma + 
C\, \hat{q}\cdot\vec{
\epsilon}_X (\hat{q}\times\vec{\epsilon}_\psi)\cdot\vec{\epsilon}_\gamma~, 
\end{eqnarray}
where $\hat{q}$ is the unit vector along the three momentum of the photon, and 
$\vec{\epsilon}_\psi$, $\vec{\epsilon}_\gamma$ and 
$\vec{\epsilon}_X$ are the corresponding polarization vectors. The 
expressions for $A$, $B$ and $C$ in terms of loop functions are given in 
Appendix~\ref{app:abc}. Because the vector charmonium produced in $e^+e^-$ 
collisions is transversely polarized, the angle between the photon momentum and 
the $\psi$ polarization vector can be related to that with respect to the beam 
axis. The relation follows from
\begin{equation}
  \overline {\sum_{\lambda=1,2}} \left| \hat q \cdot 
\vec\epsilon_\psi^{\,(\lambda)} 
\right|^2 = \frac12 \sin^2\theta_q,
\qquad\overline {\sum_{\lambda=1,2}} \left| \hat q\times\vec\epsilon_\psi 
^{\,(\lambda)} 
\right|^2 = \frac12 (1+\cos^2\theta_q),
\end{equation}
where $\theta_q$ is the angle between the photon momentum and the beam axis.
Thus, we have the angular distribution from diagrams (a,b,c) 
\begin{eqnarray}
  \frac{d\Gamma_\text{(a,b,c)}}{d\cos\theta_q}
  \al\propto\al \overline {\sum_{\lambda=1,2}} \left( 2|B|^2 \left| \hat q 
\cdot \vec\epsilon_\psi ^{\,(\lambda)} \right|^2 +  |C|^2\left| \hat 
q\times\vec\epsilon_\psi ^{\,(\lambda)} \right|^2 \right)  \propto 1 + \rho 
\cos^2\theta_q,
\end{eqnarray}
where $\rho = \left(|C|^2-2 |B|^2\right)/ \left( 2 |B|^2 + |C|^2 \right) $. For the 
$\psi(4160)\to\gamma \X$, the value is $\rho=-0.98$ so that the angular 
distribution is almost $\sim\sin^2\theta_q$.~\footnote{The value of $\rho$ 
shows that $|B|\gg|C|$, which is due to a strong cancellation between different 
loops in $C$.} 
Thus, when the long-distance part dominates the production of 
the $\X$, one may use the angular distribution to distinguish  the $P$-wave 
$D^{(*)}\bar{D}^{(*)}$ threshold and $S$-wave $D_1\bar{D}^{(*)}$ threshold 
effects. A similar idea of using angular distributions to probe the structure 
of the $\X$ was already proposed in Refs.~\cite{Mehen:2011ds,Margaryan:2013tta}.

\section{Summary}
\label{sec:sum}

In this paper, we have investigated the production of the $\X$ in the radiative
decays of  excited charmonia. These states include the $ \psi(4040)$, 
$\psi(4160)$, $ \psi(4415)$ and the $\Y$, which are the $3S$, $2D$, $4S$
charmonium and a conjectured $ D\bar D_1$ molecule, respectively. Assuming the
$\X$ is a $ D\bar D^*$ bound state, we considered its production through the
mechanism with intermediate charmed mesons. Using a NREFT, we argue that the
meson loops with all the vertices being in an $S$-wave should provide  the most 
prominent contributions. We present a power counting that is confirmed by our 
numerical studies. It  predicts that the closer to the threshold of the open 
charm intermediate states  the initial charmonium is located, the more important the 
loops are. In this context, the production rate in the decays of the $S$-wave 
charmonia $\psi(4040,4415)$, contrary to that for the $D$-wave charmonium  
$\psi(4160)$, should be small since they couple to the $ D^{(*)}\bar D_1$ in a 
$D$-wave and $D^{(*)} \bar D^{(*)}$ in a $P$-wave. The production in the $\Y$ 
decays will be strongly enhanced compared to all the other transitions studied 
in this work, if the $\Y$ is a $ D\bar D_1$ molecule, as suggested in 
Refs.~\cite{Ding:2008gr,Li:2013bca,Wang:2013cya}, since the $S$-wave coupling 
constant is  maximized in such a case. Especially, if the mechanism for the 
production of $Z_c(3900)$ in $\Y\to \pi Z_c$ proposed in 
Ref.~\cite{Wang:2013cya} is correct, the $\X$ must be copiously produced in 
$\Y\to \X \gamma$.

We also show that the measurement of the angular distribution of the radiated 
photon in $e^+e^-\to Y(4160)\to \gamma X(3872)$ should be sensitive to the 
underlying transition mechanisms.

In this study, the $\psi(4415)$ was assumed to be an $S$-wave
charmonium. However, 
if it has a
sizable mixing with a $D$-wave $ c\bar c$ component or an $S$-wave $ D^*\bar
D_1$ component (notice that it is only 10~MeV below the $D^*\bar D_1$ 
threshold), then it
can also decay into the $ \X\gamma$ through the enhanced loops with $S$-wave
couplings. Based on our calculation, we strongly suggest to search for the $\X$ 
associated with a photon in the energy region around the $\Y$ and 4.45~GeV in 
the $e^+e^-$ collisions. 

\medskip

\section*{Acknowledgments}
We are appreciate to Thomas Mehen, Eulogio Oset and Roxanne Springer for useful 
discussions and comments. This work is supported in part by the DFG and the NSFC 
through funds provided to
the Sino-German CRC 110 ``Symmetries and the Emergence of Structure in QCD'',
the EU I3HP ``Study of Strongly Interacting Matter'' under the Seventh Framework
Program of the EU, the NSFC (Grant No. 11165005 and 11035006), and the
Ministry of Science and Technology of China
(2009CB825200).


\medskip

\begin{appendix}

\renewcommand{\theequation}{\thesection.\arabic{equation}}

\section{Loop functions}
\label{app:some}
\setcounter{equation}{0}

When we neglect the widths of all the intermediate mesons, the decay amplitudes
can be expressed in the scalar three-point loop function
\begin{equation}
   I(m_1,m_2,m_3,\vec{q})
   = i\int\frac{d^4l}{(2\pi)^4} \frac{1}{\left(l^2-m_1^2+i\epsilon\right)
   \left[(P-l)^2-m_2^2+i\epsilon\right] \left[(l-q)^2-m_3^2+i\epsilon\right] },
\end{equation}
where $m_i(i=1,2,3)$ are the masses of the particles in the loop. This loop
integral is convergent. Since all the intermediate mesons in the present case
are highly nonrelativistic, the explicit expression is derived as
\begin{eqnarray}
   \label{eq:nrloop_nowidth}
   \al\al I(m_1,m_2,m_3,\vec{q}) \nonumber\\
   \al=\al \frac{-i}{8m_1m_2m_3} \int\!\frac{d^dl}{(2\pi)^d}
   \frac{1}{ \left(l^0-\frac{\vec{l}\,^2}{2m_1}+i\epsilon\right)
   \left(l^0+b_{12}+\frac{\vec{ l}\,^2}{2m_2}-i\epsilon\right)
   \left[l^0+b_{12}-b_{23}-\frac{(\vec{l}-\vec{ q})^2}{2m_3}+i\epsilon\right] }
   \nonumber\\
   \al=\al \frac{\mu_{12}\mu_{23}}{16\pi\,m_1m_2m_3} \frac{1}{\sqrt{a}} \left[
   \arctan\left(\frac{c'-c}{2\sqrt{a(c-i\epsilon)}}\right) +
   \arctan\left(\frac{2a+c-c'}{2\sqrt{a(c'-a-i\epsilon)}}\right) \right],
\end{eqnarray}
where $\mu_{ij}=m_im_j/(m_i+m_j)$ are the reduced masses, $b_{12} = m_1+m_2-M$,
$b_{23}=m_2+m_3+q^0-M$ with $M$ the mass of the initial particle, and
\begin{equation}
\label{eq:abc}
a = \left(\frac{\mu_{23}}{m_3}\right)^2 \vec{ q}\,^2, \quad c =
2\mu_{12}b_{12}, \quad c'=2\mu_{23}b_{23}+\frac{\mu_{23}}{m_3}\vec{ q}\,^2.
\end{equation}
For more information about the loop function, we refer to
Refs.~\cite{Guo:2010ak,Cleven:2011gp}. The two arctangent functions correspond
to the two cuts in the triangle diagram~\cite{Guo:2012tg}.

In the following, we give the expression for the loop with one of the mesons
having a finite width. By assigning a constant width $\Gamma_1$ to the meson
with a mass $m_1$, the first propagator in Eq.~\eqref{eq:nrloop_nowidth} is
modified to \[ \frac1 { l^0 - \vec{l}\,^2/(2 m_1) + i \Gamma_1/2 }. \]
Thus, the first cut of the triangle diagram involving $m_1$ will be influenced,
and the scalar loop integral becomes
\begin{eqnarray}
   \label{eq:nrloop_width}
   \al\al I(m_1,m_2,m_3,\vec{q}) \nonumber\\
   \al=\al \frac{\mu_{12}\mu_{23}}{16\pi\,m_1m_2m_3} \frac{1}{\sqrt{a}} \left[
   \arctan\left(\frac{c'-c}{2\sqrt{a(c-i\mu_{12}\Gamma_1)}}\right) +
   \arctan\left(\frac{2a+c-c'}{2\sqrt{a(c'-a-i\epsilon)}}\right) \right].
\end{eqnarray}

\section{Coefficients in the decay amplitudes}
\label{app:abc}

\begin{eqnarray}\nonumber
A \al=\al \sqrt{\frac{5}{6}}\, N\, g_4 \,x\, c_0\, E_\gamma
   \left[I\left(m_{D_1^0},m_{D^0},m_{D^{\text{*0}}},\vec q \right)
   + I\left(m_{D_1
   ^0},m_{D^{\text{*0}}},m_{D^0},\vec q \right) \right]\\\nonumber
B \al=\al \frac{4}{3} \sqrt{\frac{2}{15}}i\, N\, e\, g_3\,  x\,\vec{q}^{\,2}
\left(\beta + \frac1{m_c}\right) \left[5
   I^{(1)}\left(m_{D^0},m_{D^0},m_{D^{\text{*0}}},\vec q\,\right) 
   + 2 I^{(1)}\left(m_{D^{*0}},m_{D^{*0}},m_{D^0},
\vec q\, \right) \right]\\\nonumber
C\al=\al \frac{2}{3} \sqrt{\frac{2}{15}} i\, N\, e\, g_3\, x  \,\vec{q}^{\,2}
\left[ 5 \left(\beta - \frac1{m_c}\right)
   I^{(1)}\left(m_{D^{\text{*0}}},m_{D^0},m_{D^{*0}},\vec q\,\right) \right. 
\nonumber\\
   \al\al \left. - \left(\beta + \frac1{m_c}\right)
   I^{(1)}\left(m_{D^{\text{*0}}},m_{D^{\text{*0}}},m_{D^0},\vec q\,\right) 
\right]  
\end{eqnarray}
where $N=\sqrt{M_X M_\psi} $ accounts for the nonrelativistic normalization, 
and the expression for the vector loop integral $I^{(1)}(m_1,m_2,m_3,\vec q\,)$ 
can be found in 
Ref.~\cite{Guo:2010ak}.

\end{appendix}


\begin{thebibliography}{99}

\bibitem{Choi:2003ue}
  S.~K.~Choi {\it et al.}  [Belle Collaboration],
  Phys.\ Rev.\ Lett.\  {\bf 91} (2003) 262001
  [hep-ex/0309032].

\bibitem{Aaij:2013zoa}
  R. Aaij {\it et al.}  [LHCb Collaboration],
  Phys. Rev. Lett. {\bf 110} (2013)  222001
  [arXiv:1302.6269 [hep-ex]].

\bibitem{rujula}
A. De Rujula, H. Georgi, S.L. Glashow, Phys. Rev. Lett. 38, 317
(1977).


\bibitem{voloshin}
M.B. Voloshin, L.B. Okun, JETP Lett. 23, 333 (1976). Pisma Z.
Eksp. Teor. Fiz. 23, 369 (1976)

\bibitem{Tornqvist:2004qy}
  N.~A.~Tornqvist,
  Phys.\ Lett.\ B {\bf 590} (2004) 209
  [hep-ph/0402237].

\bibitem{swanson}
 E.~S.~Swanson,
  Phys.\ Rept.\  {\bf 429} (2006) 243
  [hep-ph/0601110].


\bibitem{Hanhart:2007yq}
  C. Hanhart, Y.~S.~Kalashnikova, A.~E.~Kudryavtsev and A.~V.~Nefediev,
  Phys.\ Rev.\ D {\bf 76} (2007) 034007
  [arXiv:0704.0605 [hep-ph]].


\bibitem{Aubert:2005rm}
  B.~Aubert {\it et al.}  [BaBar Collaboration],
  Phys.\ Rev.\ Lett.\  {\bf 95} (2005) 142001
  [hep-ex/0506081].


\bibitem{Ding:2008gr}
  G.-J.~Ding,
  Phys.\ Rev.\ D {\bf 79} (2009) 014001
  [arXiv:0809.4818 [hep-ph]].

\bibitem{Li:2013bca}
  M.-T.~Li, W.-L.~Wang, Y.-B.~Dong and Z.-Y.~Zhang,
  arXiv:1303.4140 [nucl-th].

\bibitem{Wang:2013cya}
  Q.~Wang, C.~Hanhart and Q.~Zhao,
  arXiv:1303.6355 [hep-ph].

\bibitem{Filin:2010se}
  A.~A.~Filin, A.~Romanov, V.~Baru, C.~Hanhart, Y.~.S.~Kalashnikova,
A.~E.~Kudryavtsev, U.-G.~Mei{\ss}ner and A.~V.~Nefediev,
  Phys.\ Rev.\ Lett.\  {\bf 105} (2010) 019101
  [arXiv:1004.4789 [hep-ph]].


\bibitem{Guo:2011dd}
  F.-K.~Guo and U.-G.~Mei{\ss}ner,
  Phys.\ Rev.\ D {\bf 84} (2011) 014013
  [arXiv:1102.3536 [hep-ph]].


\bibitem{Brambilla:2010cs}
  N.~Brambilla, S.~Eidelman, B.~K.~Heltsley, R.~Vogt, G.~T.~Bodwin, E.~Eichten,
A.~D.~Frawley and A.~B.~Meyer {\it et al.},
  Eur.\ Phys.\ J.\ C {\bf 71} (2011) 1534
  [arXiv:1010.5827 [hep-ph]].


\bibitem{Aubert:2004ns}
  B.~Aubert {\it et al.}  [BaBar Collaboration],
  Phys.\ Rev.\ D {\bf 71} (2005) 071103
  [hep-ex/0406022].


\bibitem{Acosta:2003zx}
  D.~Acosta {\it et al.}  [CDF Collaboration],
  Phys.\ Rev.\ Lett.\  {\bf 93} (2004) 072001
  [hep-ex/0312021].


\bibitem{Abazov:2004kp}
  V.~M.~Abazov {\it et al.}  [D0 Collaboration],
  Phys.\ Rev.\ Lett.\  {\bf 93} (2004) 162002
  [hep-ex/0405004].


\bibitem{Aaij:2011sn}
  R.~Aaij {\it et al.}  [LHCb Collaboration],
  Eur.\ Phys.\ J.\ C {\bf 72} (2012) 1972
  [arXiv:1112.5310 [hep-ex]].


\bibitem{Asner:2008nq}
  D.~M.~Asner, T.~Barnes, J.~M.~Bian, I.~I.~Bigi, N.~Brambilla, I.~R.~Boyko,
V.~Bytev and K.~T.~Chao {\it et al.},
  Int.\ J.\ Mod.\ Phys.\ A {\bf 24} (2009) S1
  [arXiv:0809.1869 [hep-ex]].

\bibitem{erics}
  E.~Braaten and J.~Stapleton,
  Phys.\ Rev.\ D {\bf 81} (2010) 014019
  [arXiv:0907.3167 [hep-ph]].

\bibitem{interplay}
 C.~Hanhart, Y.~S.~Kalashnikova and A.~V.~Nefediev,
  Eur.\ Phys.\ J.\ A {\bf 47} (2011) 101
  [arXiv:1106.1185 [hep-ph]].
 
\bibitem{Mehen:2011ds}
  T.~Mehen and R.~Springer,
  Phys.\ Rev.\ D {\bf 83} (2011) 094009
  [arXiv:1101.5175 [hep-ph]].

\bibitem{Margaryan:2013tta}
  A.~Margaryan and R.~P. Springer,
  arXiv:1304.8101 [hep-ph].


\bibitem{Fleming:2007rp}
  S.~Fleming, M.~Kusunoki, T.~Mehen and U.~van Kolck,
  Phys.\ Rev.\ D {\bf 76} (2007) 034006
  [hep-ph/0703168].


\bibitem{Hu:2005gf}
  J.~Hu and T.~Mehen,
  Phys.\ Rev.\ D {\bf 73} (2006) 054003
  [hep-ph/0511321].

\bibitem{Mehen:2011tp}
  T.~Mehen and D.~-L.~Yang,
  Phys.\ Rev.\ D {\bf 85} (2012) 014002
  [arXiv:1111.3884 [hep-ph]].
 
\bibitem{Gamermann:2009fv}
  D.~Gamermann and E.~Oset,
  Phys.\ Rev.\ D {\bf 80} (2009) 014003
  [arXiv:0905.0402 [hep-ph]].
  
\bibitem{Gamermann:2009uq}
  D.~Gamermann, J.~Nieves, E.~Oset and E.~Ruiz Arriola,
  Phys.\ Rev.\ D {\bf 81} (2010) 014029
  [arXiv:0911.4407 [hep-ph]].
 
\bibitem{Aceti:2012cb}
  F.~Aceti, R.~Molina and E.~Oset,
  Phys.\ Rev.\ D {\bf 86} (2012) 113007
  [arXiv:1207.2832 [hep-ph]].

\bibitem{Guo:2009wr}
  F.-K.~Guo, C.~Hanhart and U.-G.~Mei{\ss}ner,
  Phys.\ Rev.\ Lett.\  {\bf 103} (2009) 082003
   [Erratum-ibid.\  {\bf 104} (2010) 109901]
  [arXiv:0907.0521 [hep-ph]].


\bibitem{Guo:2010zk}
  F.-K.~Guo, C.~Hanhart, G.~Li, U.-G.~Mei{\ss}ner and Q.~Zhao,
  Phys.\ Rev.\ D {\bf 82} (2010) 034025
  [arXiv:1002.2712 [hep-ph]].


\bibitem{Guo:2010ak}
  F.-K.~Guo, C.~Hanhart, G.~Li, U.-G.~Mei{\ss}ner and Q.~Zhao,
  Phys.\ Rev.\ D {\bf 83} (2011) 034013
  [arXiv:1008.3632 [hep-ph]].


\bibitem{Guo:2012tg}
  F.-K.~Guo and U.-G.~Mei{\ss}ner,
  Phys.\ Rev.\ Lett.\  {\bf 109} (2012) 062001
  [arXiv:1203.1116 [hep-ph]].


\bibitem{Beringer:1900zz}
  J.~Beringer {\it et al.}  [Particle Data Group],
  Phys.\ Rev.\ D {\bf 86} (2012) 010001.


\bibitem{Thomas:2008ja}
  C.~E.~Thomas and F.~E.~Close,
  Phys.\ Rev.\ D {\bf 78} (2008) 034007
  [arXiv:0805.3653 [hep-ph]].


\bibitem{Fleming:2008yn}
  S.~Fleming and T.~Mehen,
  Phys.\ Rev.\ D {\bf 78} (2008) 094019
  [arXiv:0807.2674 [hep-ph]].


\bibitem{Godfrey:1985xj}
  S.~Godfrey and N.~Isgur,
  Phys.\ Rev.\ D {\bf 32} (1985) 189.


\bibitem{Barnes:2005pb}
  T.~Barnes, S.~Godfrey and E.~S.~Swanson,
  Phys.\ Rev.\ D {\bf 72} (2005) 054026
  [hep-ph/0505002].


\bibitem{Amundson:1992yp}
  J.~F.~Amundson, C.~G.~Boyd, E.~E.~Jenkins, M.~E.~Luke, A.~V.~Manohar,
J.~L.~Rosner, M.~J.~Savage and M.~B.~Wise,
  Phys.\ Lett.\ B {\bf 296} (1992) 415
  [hep-ph/9209241].


\bibitem{Weinberg:1965zz}
  S.~Weinberg,
  Phys.\ Rev.\  {\bf 137} (1965) B672.


\bibitem{Baru:2003qq}
  V.~Baru, J.~Haidenbauer, C.~Hanhart, Y.~S.~Kalashnikova and A.~E.~Kudryavtsev,
  Phys.\ Lett.\ B {\bf 586} (2004) 53
  [hep-ph/0308129].


\bibitem{Cleven:2011gp}
  M.~Cleven, F.-K.~Guo, C.~Hanhart and U.-G.~Mei{\ss}ner,
  Eur.\ Phys.\ J.\ A {\bf 47} (2011) 120
  [arXiv:1107.0254 [hep-ph]].


\bibitem{Fayyazuddin:1994qu}
  Fayyazuddin and O.~H.~Mobarek,
  Phys.\ Rev.\ D {\bf 50} (1994) 2329.


\bibitem{Korner:1992pz}
  J.~G.~Korner, D.~Pirjol and K.~Schilcher,
  Phys.\ Rev.\ D {\bf 47} (1993) 3955
  [hep-ph/9212220].


\bibitem{Godfrey:2005ww}
  S.~Godfrey,
  Phys.\ Rev.\ D {\bf 72} (2005) 054029
  [hep-ph/0508078].

\bibitem{Badalian:2008dv}
  A.~M.~Badalian, B.~L.~G.~Bakker and I.~V.~Danilkin,
  Phys.\ Atom.\ Nucl.\  {\bf 72} (2009) 638
  [arXiv:0805.2291 [hep-ph]].
  
\bibitem{Aubert:2008ae}
  B.~Aubert {\it et al.}  [BaBar Collaboration],
  Phys.\ Rev.\ Lett.\  {\bf 102} (2009) 132001
  [arXiv:0809.0042 [hep-ex]].  
  
\bibitem{Bhardwaj:2011dj}
  V.~Bhardwaj {\it et al.}  [Belle Collaboration],
  Phys.\ Rev.\ Lett.\  {\bf 107} (2011) 091803
  [arXiv:1105.0177 [hep-ex]].
  
\bibitem{Ablikim:2013mio}
  M.~Ablikim {\it et al.}  [ BESIII Collaboration],
  Phys.\ Rev.\ Lett.\  {\bf 110} (2013) 252001
  [arXiv:1303.5949 [hep-ex]].

\bibitem{Liu:2013dau}
  Z.~Q.~Liu {\it et al.}  [Belle Collaboration],
  Phys.\  Rev.\  Lett.\ {\bf 110} (2013) 252002
  [arXiv:1304.0121 [hep-ex]].

\bibitem{Xiao:2013iha}
  T.~Xiao, S.~Dobbs, A.~Tomaradze and K.~K.~Seth,
  arXiv:1304.3036 [hep-ex].

\bibitem{Wang:2013hga}
  Q.~Wang, C.~Hanhart and Q.~Zhao,
  arXiv:1305.1997 [hep-ph].

\bibitem{Guo:2013sya}
  F.-K.~Guo, C.~Hidalgo-Duque, J.~Nieves and M.~P.~Valderrama,
  arXiv:1303.6608 [hep-ph].

\bibitem{Cui:2013yva}
  C.-Y.~Cui, Y.-L.~Liu, W.-B.~Chen and M.-Q.~Huang,
  arXiv:1304.1850 [hep-ph].

\bibitem{Wilbring:2013cha}
  E.~Wilbring, H.-W.~Hammer and U.-G.~Mei{\ss}ner,
  arXiv:1304.2882 [hep-ph].
  
\bibitem{Zhang:2013aoa}
  J.-R.~Zhang,
  Phys.\  Rev.\  D 87, {\bf 116004} (2013)
  [arXiv:1304.5748 [hep-ph]].

\bibitem{Liu:2013}
  X.-H.~Liu and G.~Li,
  arXiv:1306.1384 [hep-ph].

\end{thebibliography}

\end{document}